\documentclass[aps,prb,twocolumn,showpacs,superscriptaddress]{revtex4}
\usepackage{graphicx}
\usepackage{verbatim}
\usepackage{amsmath}
\usepackage{sidecap}
\usepackage{epstopdf}

\begin{document}

\title{Multiplet resonance lifetimes in resonant inelastic X-ray scattering involving shallow core levels}

%\title{The importance of self consistent lifetime correction for resonant X-ray scattering involving shallow core levels}
%\title{Resonant inelastic X-ray scattering from model Mott insulators matches a lifetime corrected first principles-based model}
% \title{Quantitatively accurate simulation of the $dd$ resonant inelastic X-ray scattering spectrum for nickel oxide}

\author{L. Andrew Wray}
\author{Wanli Yang}
\affiliation{Advanced Light Source, Lawrence Berkeley National Laboratory, Berkeley, California 94305, USA}
\author{Hiroshi Eisaki}
\affiliation{Nanoelectronic Research Institute, National Institute of Advanced Industrial Science and Technology, Tsukuba, 305-8568, Japan}
\author{Zahid Hussain}
\author{Yi-De Chuang}
\affiliation{Advanced Light Source, Lawrence Berkeley National Laboratory, Berkeley, California 94305, USA}


\begin{abstract}

Resonant inelastic X-ray scattering (RIXS) spectra of model copper- and nickel-based transition metal oxides are measured over a wide range of energies near the M-edge (h$\nu$=60-80eV) to better understand the properties of resonant scattering involving shallow core levels. Standard multiplet RIXS calculations are found to deviate significantly from the observed spectra. However, by incorporating the self consistently calculated decay lifetime for each intermediate resonance state within a given resonance edge, we obtain dramatically improved agreement between data and theory. Our results suggest that these textured lifetime corrections can enable a quantitative correspondence between first principles predictions and RIXS data on model multiplet systems. This accurate model is also used to analyze resonant elastic scattering, which displays the elastic Fano effect and provides a rough upper bound for the core hole shake-up response time.

\end{abstract}

% \pacs{}

\date{\today}

\maketitle

\section{Introduction}

An accurate understanding of resonant X-ray interactions with matter is of central importance for current investigations into low energy many-body structure and dynamics with resonant elastic (REXS) and resonant inelastic scattering (RIXS). Atomic multiplet calculations have provided the central basis for interpreting the so called ``direct" scattering processes (non-shake-up processes) \cite{deGrootEarlyNiO,KotaniIdea,KotaniRIXSreview,AmentRIXSReview} by expanding upon a very precise formulation of intra-atomic electronic interactions \cite{RacahMultFormula}. Although these models have achieved remarkable success, current understanding of how they can be best implemented to study low energy material properties is limited by the complexity of many-body systems. There are few model systems for which close correspondence can be found between first principles-based predictions and experimental RIXS spectra across the full incident energy range of resonance, particularly if one excludes the frequently studied cuprate compounds.

\begin{figure}[t]
\includegraphics[width = 8cm]{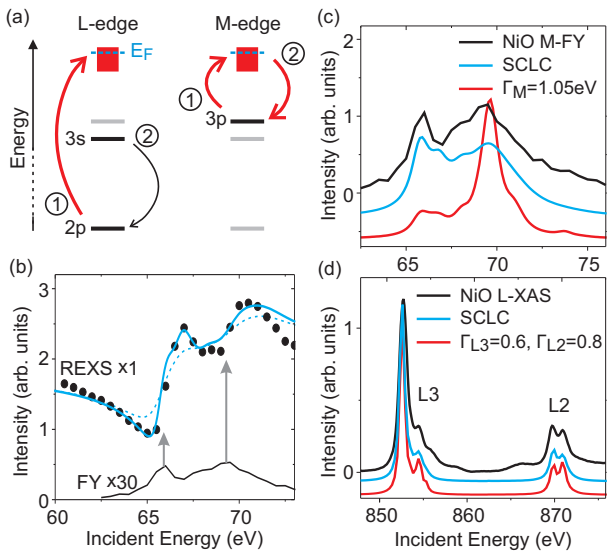}
\caption{{\bf{Shallow vs. deep core level resonant decay}}: (a) A diagram shows resonant excitation at the M- and L-edges of 3d transition metals, and the lowest energy dipole-allowed decay channel of intermediate states. Red lines indicate scattering processes that interact with electrons near the Fermi level. (b) (`REXS', dots) Integrated elastic and (`FY') inelastic scattering intensities measured at the NiO M-edge are compared on an absolute, unshifted scale. Blue curves show (solid) the SCLC REXS simulation and (dashes) a modified simulation with a strong competing decay channel ($C_{CC}$=1.0eV). (c) Fluorescence yield measured at the M-edge of NiO is compared with vertically shifted curves showing (red) a standard multiplet calculation with inverse lifetime $\Gamma_m$=1.05eV, and (blue) the same model using SCLC. (d) (black) Total electron yield (representing X-ray absorption) measured at the L-edge is compared to (red) a standard multiplet calculation with constant intermediate state inverse lifetimes $\Gamma_{L3}$=0.6eV and $\Gamma_{L2}$=0.8eV, and (blue) the same calculation with SCLC valence transitions of $C_V(0.3eV)$ and $C_{CC}$=0.2eV at L$_3$. $C_{CC}$ is increased to 0.5eV at L$_2$ due to Coster-Kronig decay \cite{L2vsL3lifetime}, but $C_V$ remains constant.}
\end{figure}

In this article, we investigate RIXS spectra at the soft X-ray M$_2$-M$_3$ edges of SrCuO$_2$ and NiO, two relatively simple transition metal oxide systems that are well suited to analysis by atomic multiplet simulations. Our principle finding is that, for measurements on model Mott insulating systems, quantitative correspondence can be achieved between RIXS data and first principles based predictions by incorporating self-consistent lifetime corrections (SCLC) into the framework of existing models. Atomic multiplet interactions are found to lead to more than 10 distinct lifetimes ranging from $\sim$1-4eV for different intermediate states at the Ni M-edge. Our results show that, for the studied compounds, these variations in lifetime can be properly accounted for by low order SCLC to dramatically improve the accuracy of resonant scattering simulations. Similar lifetime calculations have been performed to explain feature widths in measurements of core electronic states \cite{KotaniIdea,simCalc}, however we will show that SCLC has more far reaching significance for the low energy degrees of freedom observed with RIXS and REXS.

Nickel oxide is a model spin 1 antiferromagnet, with Ni at d8 valency. Low energy excitations in NiO M-edge RIXS spectra have been shown to correspond well to multiplet predictions \cite{deGrootEarlyNiO,NiO_MRIXS,NiO_LRIXS,J0p1eV,MEcalc}. SrCuO$_2$, with Cu at d9 and d10 valencies in the ground and intermediate states \cite{ZXOrbitals}, is an easily solvable model system from the standpoint of multiplet physics \cite{WrayArXiV}. Comparison between these two materials not only provides an opportunity to test the validity of current atomic multiplet models in describing RIXS spectra involving extremely shallow core levels, but also offers a unique perspective into the detailed RIXS process at such low excitation energies.

M-edge measurements were performed at the beamline 4.0.3 (MERLIN) RIXS endstation (MERIXS) at the Advanced Light Source (ALS), Lawrence Berkeley National Laboratory. The data were recorded by a VLS based X-ray emission spectrograph equipped with a commercially available CCD detector \cite{YiDeDetector}. Scattering intensity measured at each pixel was normalized to ensure that intensity in the final spectra accurately represents the density of scattered photons per unit energy. Large cleaved single crystal samples were measured at a pressure of 3$\times$10$^{-10}$ Torr at room temperature. The resolution-limited full width at half maximum of the elastic line is better than $\delta$E$\lesssim$25$\pm$2 meV for NiO M-edge measurements, and $\delta$E$\sim$200meV for the L-edge X-ray electron yield (XAS). Since the proximity of the [000] Bragg peak at the M-edge creates tremendous intensity at the elastic line (see Fig. 1(b)), making it challenging to measure fluorescence yield (FY) in a conventional way, FY in this study is measured by integrating RIXS intensity within a 0.8-3.4eV energy loss window. Measurements at the M-edge of NiO were performed with in-plane polarization in the [001] scattering plane, and the photon beam had a grazing 25$^o$ angle of incidence to the cleaved [100] face. L-edge measurements were performed in the same configuration with a 45$^o$ angle of photon incidence. Charging in was minimized by mounting a thin, flat NiO crystal on conducting carbon tape, and wrapping the tape around the edges of the sample. Adjusting photon flux with the upstream slits did not noticeably change the XAS spectrum. All simulations are performed with photon geometries matching the experiment, and energy loss features are convoluted by a 100meV Lorentzian function.

\section{Lifetime effects in fluorescence spectra}

%Correspondence between incident energy dependence in RIXS data and first principles-based simulations is often weak for materials with strong many-body dynamics, and it can be difficult to experimentally identify the factors that are important for accurate modeling. The shallow 3p core level resonance modes selected for this study provide extremely weak self absorption cross sections \cite{XrayPath} and strong Coulomb interactions with valence 3d orbitals, both positive factors for simulating RIXS with an atomic multiplet basis.

The significance self consistently calculated intermediate state lifetimes can have for resonant inelastic scattering is evident in the stark contrast between M-edge NiO fluorescence yield spectra simulated with and without SCLC in a standard atomic multiplet model (Fig. 1(c), modeling details in Section III). The difference between corrected and uncorrected models becomes smaller in the harder X-ray regime, where lifetime is increasingly defined by core hole decay channels such as the L-edge 3s-2p Coster-Kronig transition \cite{L2vsL3lifetime,Kotani3sDecay} that do not depend strongly on valence electron symmetries (Fig. 1(a)). Also, core hole shake-up excitations are relatively strong at the L-edge \cite{NiO_MRIXS,CoO_MRIXS,WrayArXiV}, and have a nonlinear time dependence \cite{coreClock,Abbamonte3rdOrder,VDBultrashort} that helps unify the lifetimes of multiplet RIXS features. Nonetheless, SCLC at the L$_{2,3}$-edges appears to somewhat improve the correspondence between theory and data for the measurement geometry in Fig. 1(d) by making the higher energy L$_2$ peak broader and less prominent.

The resonant elastic spectrum in Fig. 1(b) provides another perspective from which to observe the accuracy of the SCLC lifetimes. Elastic scattering is measured by integrating intensity on the RIXS analyzer from -0.1 to 0.1 eV energy loss. Due to the low photon energies used in this study, scattering in this energy loss window is dominated by the tail of the [000] Bragg peak, and features strong elastic Fano interference \cite{elastFano} between resonant and non-resonant scattering channels. This interference creates a Fano dip at the leading edge of resonant intensity, and causes resonant elastic features to be shifted to higher energy relative to inelastic resonance peaks (indicated by arrows). We note that the origin of this Fano effect is similar to the Auger Fano effect \cite{OgasawaraFano}, but differs in that the strength of interference in this case depends on rocking curves of the [000] Bragg peak. Fano interference in the calculation is obtained by estimating a frequency-independent cross section for non-resonant scattering. A gentle slope in the data from decay of [000] Bragg intensity is not factored into the calculation, and causes the simulation intensity to be too weak at low energy (small Q) and too strong at the high energy (large Q) edge of the spectrum.

Low energy core hole shake-up responses of the many-body system, including the creation of phonons or many-body spin excitations, will not change the lifetimes observed in inelastic emission because they do not cause the core hole to decay, but can shorten REXS lifetimes by forbidding relaxation to the ground state. The dashed curve in Fig. 1(b) shows how the simulated REXS lineshapes change when a competing decay channel with inverse lifetime $C_{CC}$=1.0eV is factored in to represent low energy many-body response (see Equation (4)). The uncorrected SCLC lineshape provides a qualitatively better match for data, including the curvature of the Fano dip, suggesting that the low energy collective response is slow relative to the SCLC calculated core hole decay rates of $\Gamma_m$=1-4eV.

\begin{SCfigure*}
\centering
\includegraphics[width = 11cm]{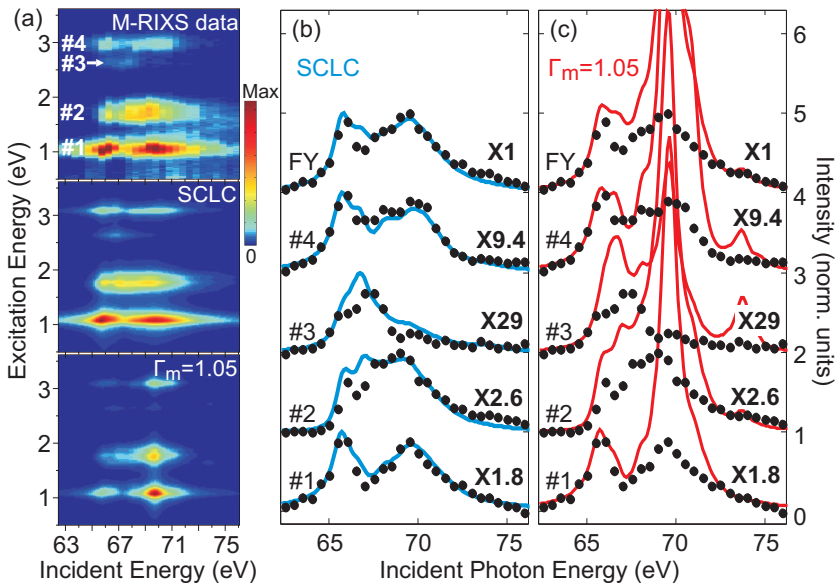}
\caption{{\bf{Lifetime corrected simulation of M-edge scattering}}: (a) (top) Observed excitation peaks are numbered $\#$1-4 on a plot of the incident energy dependence of M-edge RIXS scattering from NiO. A multiplet simulation is shown (middle) with SCLC and (bottom) with constant $\Gamma_m$=1.05eV. (b-c) Spectral intensity is integrated from (dots) the RIXS data and (lines) simulations for excitation energy windows of ($\#$1) 0.8-1.3eV, ($\#$2) 1.35-2.2eV, ($\#$3) 2.3-2.65eV, ($\#$4) 2.75-3.25eV, (FY) 0.8-3.4eV. These correspond roughly to the 4 excitations predicted by multiplet theory, and to the inelastic fluorescence yield. The experiment and simulations are scaled by the same factor indicated next to each curve. Statistical error from photon count and detector dark current is insignificant.}
\end{SCfigure*}

\section{Direct RIXS scattering and SCLC}

The direct scattering process of RIXS and REXS is described by the Kramers-Heisenberg equation as follows:

\begin{align}
    R_{f}(E,h\nu,q) \propto \sum_{q',g}\left|\sum_m\frac{
    \langle f|T^{\dagger}_{q'}| m\rangle
    \langle m|T_q|g\rangle}{h\nu-E_m+i\Gamma_m/2}
    \right|^2 \nonumber \\
    \times\frac{\frac{1}{2\pi}\Gamma_{f}}{(E-E_{f})^2+(\frac{1}{2}\Gamma_{f})^2}
\end{align}

where $g$ sums over the degenerate ground states, and $m$ runs over intermediate states where a core hole is present. Incident photon energy is written as $h\nu$, the excitation energy is $E$, and the inverse final state ($|f>$) excitation lifetime is $\Gamma_{f}$. The photon perturbation $T_q$ is given in the dipole approximation by a sum over spherical harmonics with $q=-1,0,1$ for \emph{left}, \emph{linear} ($z$) and \emph{right} polarized light. The sum over $g$ and geometry-dependent sum over $q'$ are required because the measurements are performed over a $\sim$0.5mm long strip on the sample encompassing many magnetic and structural domains \cite{NatPhotNiO}, with no discrimination of the polarization of scattered photons. Unlike the L-edges, the M$_2$ and M$_3$ edges are merged together and not separately visible for late transition metal oxides other than d9 systems such as cuprates \cite{NiO_MRIXS,CoO_MRIXS,MEcalc,WrayArXiV}.

We have modified the equation from its most common presentation by explicitly indexing the lifetime $\Gamma_m$ of each intermediate state. Aside from this, our treatment follows that in Ref. \cite{NiO_MRIXS}. Hartree-Fock values are reduced to 80$\%$ and 70$\%$ for the inter-d-orbital and 3p-3d Slater-Condon parameters, and the d-orbital spin orbit coupling parameter is set to 0.083eV \cite{SlaterCondonParams}. The intrinsic spin orbit splitting between 3p$_{1/2}$ and 3p$_{3/2}$ is set to 2.35eV, inflated by 30$\%$ from the nominal value of 1.8eV for pure nickel due to reduced screening with positive 4s valence \cite{BE3,WrayArXiV}. All states are solved for by exact diagonalization on an atomic multiplet state basis. The crystal field perturbation of 10Dq=1.03eV is fixed by the RIXS excitation energies as considered in Ref. \cite{NiO_MRIXS}. An external exchange field of $J^*=0.1eV$ is applied to the e$_g$ orbitals to account for oxygen-mediated large spin antiferromagnetism \cite{J0p1eV}. The exchange field is summed over all 12 [11$\overline{2}$] magnetic domain orientations; however, the key features discussed in this paper do not change significantly when $J^*$ is set to zero (as in Fig. 3(a, bottom)).

\begin{figure}[t]
\includegraphics[width = 8cm]{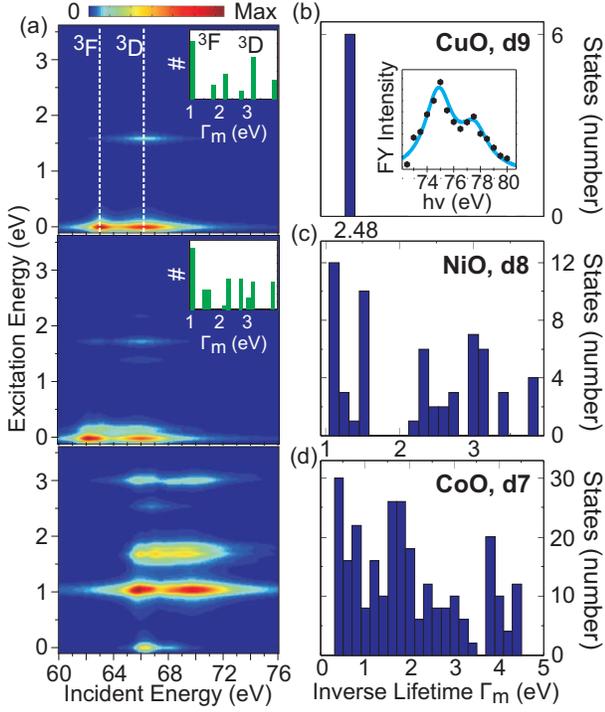}
\caption{{\bf{The origin of M-edge resonant lifetimes}}: (a, top) Simulated SCLC RIXS scattering for an isolated Ni$^{2+}$ ion with spin-orbit coupling disabled. Dipole excited intermediate state symmetries are highlighted. (a, middle) Simulated SCLC RIXS scattering from an isolated Ni$^{2+}$ ion. (a, bottom) Simulated SCLC RIXS scattering from Ni$^{2+}$ perturbed by an octahedral crystal field. Insets show histograms of the intermediate state inverse lifetimes. (b-d) Histograms show intermediate state lifetime distributions calculated for cuprate, nickelate and cobaltate M$_{2/3}$ RIXS scattering, using the cross section parameters $C_{3d,3p}$ and $C_{CC}$ fitted for NiO in all three cases. The inset of panel (b) shows that the M$_3$ and M$_2$ features of SrCuO$_2$ are accurately fitted with the predicted $\Gamma$=2.5eV lifetime.}
\end{figure}

For the RIXS simulation in Fig. 2(a,bottom), the inverse lifetime $\Gamma_m$ is set to a fixed value for all intermediate states ($\Gamma_m$=1.05eV), as in previous studies. Variants of this fixed-lifetime model have been used with considerable success at the leading edge of M-resonance \cite{NiO_MRIXS,CoO_MRIXS}, and for the L-edge with distinct lifetime parameters $\Gamma$ for L$_2$ and L$_3$ resonance \cite{L2vsL3lifetime,deGrootEarlyNiO,NiO_LRIXS,J0p1eV}. When compared across the full incident energy range of M$_{2/3}$ resonance however, the deviation from theory is striking. Although inelastic features labeled $\#$1-4 are found at approximately correct energies, their predicted spectral intensities correspond very poorly with the experimental data in Fig. 2(a, top).

Integrated feature intensity is shown as a function of incident energy in Fig. 2(b-c), and is likely to be the best basis for rigorous comparison between the simulations and data, because the integration window is large enough that final states spread out on a $\sim$0.1eV scale by many-body effects will be fully counted. Features at higher incident energies are clearly too intense in the $\Gamma_m$=1.05eV simulation and have lineshapes that are too sharp with respect to the data. Even at lower resonance energies near 66eV, the simulated intensity of features $\#$2-3 is too strong relative to feature $\#$1. To improve upon this modeling paradigm, we note that the inverse lifetime of an intermediate state is derived from the allowed decay channels, described to lowest order via Fermi's golden rule \cite{degenStates}:
%GrifParticle
\begin{align}
    \Gamma_m = C \sum_{q',f}\left|\langle f|T^{\dagger}_{q'}| m\rangle\right|^2
\end{align}

where C is a constant. These matrix elements are obtained in the course of solving the Kramers-Heisenberg equation in multiplet RIXS models, to within a common multiplicative factor ($C_{ij}$) that depends on the radial wavefunctions of electron shells between which a transition occurs:

\begin{align}
    \Gamma_m = \sum_{q',f,i,j}C_{i,j}\left|\langle f_{i,j}|T^{\dagger}_{q'}| m\rangle\right|^2
\end{align}

For NiO resonant scattering at the M-edge, $f_{3d,3p}$ indexes a final state in which the intermediate state has been terminated by the transition of a 3d electron into the 3p core level, and at the L-edge, $f_{3s,2p}$ would represent a final state following from Koster-Cronig transition. Because decay channels that do not involve valence electrons give roughly constant contributions to the lifetime, it is convenient to reframe this as:

\begin{align}
    \Gamma_m = C_{CC}+\sum_{q',f_V}C_V(\Gamma_{min}-C_{CC})\left|\langle f_{V}|T^{\dagger}_{q'}| m\rangle\right|^2
\end{align}

where $C_{CC}$ is a constant representing the sum over transitions between core levels (and any other competing decay channels), f$_V$ indexes all final states of valence electron decay, and the valence decay rate constant $C_V(\Gamma_{min}-C_{CC})$ is defined such that the longest lived intermediate state has an inverse lifetime of $\Gamma_{min}$. In the simulation of the L$_2$ edge (Fig. 1(d)), using the values $C_V(0.3eV)$ and $C_{CC}$=0.5eV causes the smallest intermediate state lifetime to be $\Gamma_{min}$=0.3+0.5=0.8eV. To incorporate this state-by-state determination of intermediate state widths in our simulation for NiO M-edge RIXS, we use values of $C_V(1.0eV)$ and $C_{CC}$=0.05. This assumes that competing decay channels not considered by the multiplet calculation, such as indirect RIXS excitations \cite{indRIXSdeq} and direct transitions from oxygen p-orbitals to the nickel 3p level, are very slow and constitute only 0.05eV out of the sum. The combined value of $\Gamma_{min}$=1.05eV ($\sim$1eV) is chosen to match the sharpest lineshapes in the spectra, and thus corresponds closely to the value of $\Gamma$=1eV fitted in earlier fixed-lifetime studies \cite{earlyLifetime,NiO_MRIXS}.

%%% SCLC and RIXS

Repeating the Kramers-Heisenberg multiplet RIXS calculation for NiO with intermediate state lifetimes determined in this self consistent way gives the SCLC spectrum shown in Fig. 2(a, middle), which is in far better agreement with the experimental data. Curves tracing the intensity of inelastic features in the data and simulation show only minor discrepancies (Fig. 2(b)), and the calculation very accurately reproduces the relative intensities of all features. The SCLC lifetimes do not just broaden the data, but also act as a physically necessary normalization factor of Equation (1). Without this normalization, the higher energy resonance around 70eV becomes far too intense, as well as being too sharp (Fig. 2(c)). Line shapes of resonance are generally broader at higher incident energies of h$\nu$$\sim$70eV, and calculated intermediate state lifetimes become shorter in that region. Even at fixed incident energy near the leading edge of resonance (h$\nu$$\sim$66eV), the relative intensities of features $\#$1-4 are improved by SCLC, because of the selective coupling between different intermediate and final state symmetries. It is likely that the agreement between calculations and experiments can be further improved by including other energetic parameters such as the difference in nephelauxetic renormalization of Slater-Condon terms for e$_g$ and t$_{2g}$ orbitals. However, we suggest that caution is necessary if one wishes to explicitly include oxygen hybridization, because the common practice of discarding the interatomic Slater-Condon terms between near neighbor copper and oxygen orbitals will create inaccuracy when calculating intermediate state lifetimes.

\section {Discussion}

Fig 3(a) illustrates how the diversity in M-edge intermediate state lifetimes comes about. For an isolated ion with no spin orbit interactions, few excitations are dipole allowed. However, intermediate state lifetimes are not single valued (inset in Fig. 3(a, top)). As an example, the $^3$F intermediate state with angular momentum moment m$_L$=+3 (i.e. L=3, m$_L$=+3) is composed of a 3p core hole with m$_L$=+1 and a 3d hole with m$_L$=+2. Selection rules dictate that the 3p core hole can be filled by d-electrons with m$_L$=0 or +2, but there is a hole in the m$_L$=+2 3d-orbital, so only the m$_L$=0 d-electron can transition to terminate the intermediate state. Thus, even without considering the detailed dipole matrix elements, one can see that the $^3$F intermediate state will tend to take longer to decay than other intermediate states in which the p- and d-holes have opposed angular momentum.

The addition of spin orbit coupling to the simulated ion causes new excitations to become allowed, but does not change the spectrum dramatically (Fig. 3(a, middle)). However, by breaking continuous rotational symmetry with an octahedral crystal field (10Dq=1.03eV as for NiO), numerous non-zero transition matrix elements are created (Fig. 3(a, bottom)). All features are shifted to higher energy by the crystal field perturbation, yet a remnant correspondence with the narrow $^3$F and broad $^3$D intermediate state lineshapes is visible in narrow features at $\sim$66eV and broader features at $\sim$70eV, respectively. This panel already closely resembles the experimental data, in spite of not including the antiferromagnetic exchange field.

Another useful point of comparison is M-edge scattering from SrCuO$_2$, which has full d10 intermediate state valence and thus only one intermediate state lifetime. Remarkably, using the parameters $C_{3d,3p}$ and $C_{CC}$ derived from NiO results in an inverse lifetime of $\Gamma_m$=2.48eV if starting valence is changed to d9 as for cuprates. This value correctly fits M-edge FY measured on SrCuO$_2$ (Fig. 3(b, inset)), and is within error bars of the measured SrCuO$_2$ core hole lifetime value of $\Gamma$=2.5$\pm$0.3eV \cite{WrayArXiV}. The fact that the core hole lifetime of SrCuO$_2$ can also be accurately predicted from the multiplet calculation using Equations (3-4) gives an extremely strong indication that core hole decay processes at the M-edge are accurately described by the model after inclusion of SCLC. Histograms of intermediate state lifetimes determined by multiplet calculation for NiO and CoO are presented in Fig. 3(c-d) showing that shifting the valence state away from full d-shell occupation increases the range of intermediate state lifetimes, and a lifetime-corrected model is likely to be even more important for 3d transition metal compounds with close to half-filled 3d shells.

Resonant scattering involving shallow core levels is not as well studied as harder X-ray resonance, largely because of the weaker X-ray emission signal from strong Auger decay and poor penetration skin depths for soft X-rays. It is worth noting however that resonant self absorption is weaker relative to the non-resonant absorption background at low photon energies, making close comparison between X-ray emission data and theory more practical \cite{WrayArXiV,XrayPath}. The corrections we identify are also relevant to harder X-ray regimes, and their effects in RIXS, REXS and FY scattering are enhanced by the need for accurate lifetime values to normalize the scattering equation (Eq. (1)) and identify the onset of core hole shake-up excitations \cite{AmentRIXSReview,VDBultrashort}. Significant quantum interference between atomic multiplet scattering channels is also moderated by the intermediate state lifetimes \cite{WrayArXiV}. These and other factors can generate tremendous complexity in theories of resonant scattering, thus it is essential to characterize model systems such as M-edge NiO to identify the underlying scattering physics.

In conclusion, through comparison of M-edge resonant inelastic X-ray scattering data on model systems NiO and SrCuO$_2$ with an atomic multiplet model, we have shown that intermediate state lifetimes for M-edge resonant inelastic scattering can be extremely diverse over a small incident energy range. By applying self consistent lifetime corrections (SCLC), the full inelastic resonance profile of NiO is simulated with greatly enhanced accuracy. Consistent lifetime-derived lineshapes on the elastic line elastic line show that the core hole shake-up response time is slow relative to core hole decay. Evaluating the case of CoO demonstrates that the same principle can be expected to apply for resonant scattering on other compounds with shallow core levels, and determination of these lifetime factors is critical for modeling resonant soft X-ray probes of low energy many-body states.

\textbf{Acknowledgements:}

L.A.W. acknowledges discussions with Elke Arenholz. The excellent CTM4XAS code maintained by F.M.F. de Groot at http://www.anorg.chem.uu.nl/CTM4XAS/ was used to test our multiplet diagonalization calculations, and as a source of Slater-Condon parameters. The synchrotron X-ray-based measurements and theoretical computations are supported by the Basic Energy Sciences of the US DOE (DE-FG-02-05ER46200, AC03-76SF00098 and DE-FG02-07ER46352).


\begin{thebibliography}
\bibitem{degenStates} Degenerate states are assigned an average lifetime value, to ensure basis independence.
\bibitem{indRIXSdeq} Strong indirect RIXS processes do not directly effect broadening for XAS, but will influence the functional definition of $\Gamma_m$ for RIXS with sufficiently long core hole lifetimes (e.g. L-edge).
\end{thebibliography}
\end{document}